\documentclass[conference]{IEEEtran}
\IEEEoverridecommandlockouts
\usepackage{cite}
\usepackage{amsmath,amssymb,amsfonts}
\usepackage{algorithmic}
\usepackage{graphicx}
\usepackage{textcomp}
\usepackage{xcolor}
\usepackage{amsthm}
\usepackage{verbatim}

\def\BibTeX{{\rm B\kern-.05em{\sc i\kern-.025em b}\kern-.08em
    T\kern-.1667em\lower.7ex\hbox{E}\kern-.125emX}}
\begin{document}

\title{A Public Safety Framework for Immersive Aerial Monitoring through 5G Commercial Network\\
\thanks{Sejin Seo and Seunghwan Kim are co-first authors. This work was supported by Institute of Information \& communications Technology Planning \& Evaluation (IITP) grant funded by the Korea government (MSIT) (No.2018-0-00170, Virtual Presence in Moving Objects through 5G).}
}

\author{\IEEEauthorblockN{ Sejin Seo, Seunghwan Kim, and Seong-Lyun Kim} \\
\IEEEauthorblockA{School of Electrical and Electronic Engineering, Yonsei University\\
50 Yonsei-Ro, Seodaemun-Gu, Seoul 03722, Korea\\
Email: \{sjseo, shkim, slkim\}@ramo.yonsei.ac.kr}}

\maketitle

\begin{abstract}

Are 5G connection and UAVs merely parts of an extravagant and luxurious world, or are they essential parts of a practical world in a way we have yet to see? To aid in a direction to address the issue, we provide a practical framework for immersive aerial monitoring for public safety. Because the framework is built on top of actual realizations and implementations designed to fulfill specific use cases, high level of practicality is ensured by nature. We first investigate 5G network performance on UAVs by isolating performance for different aspects of expected flight missions. Finally, the novel aerial monitoring scheme that we introduce relies on the recent advances brought by 5G networks and mitigates the inherent limitations of 5G network that we investigate in this paper.
\end{abstract}


\section{Introduction}
Immersive media, e.g. virtual reality (VR) and augmented reality (AR), gifts users immersive remote presence in a virtual world or physically distant locations. While the majority of immersive contents are currently consumed for video games and more broadly to entertain, its usage outside of entertainment deserves more attention. Apart from use cases regarding entertainment services, it is still disputable, thus needing more research on whether the extra immersiveness brought by immersive media improves the key quality of the associated activity, as in quality of learning during a science lab \cite{b1}. Nevertheless, real-time immersive aerial monitoring is a promising way to improve public safety, e.g. in fire fighting scenario of \cite{b2}. In addition, aerial monitoring using low cost UAVs is presented to improve the resolution of agricultural remote sensing \cite{b3}. In consequence, in this paper, we focus on designing a system, which supports immersive media, to be actually helpful, instead of merely adding flair to the system, which could end up as a distraction and overhead, given the gravity of the use cases we are concerned with. 

Notwithstanding the public regulations, aerial vehicles, and more specifically unmanned aerial vehicles (UAVs), provide extended flexibility in mobility, diversity of sensing, and reduction of mission latency. Extended flexibility enables the drones to traverse to locations that are necessary to be covered but unreachable to terrestrial fire fighting vehicles. By reaching different locations in 3 dimensions, UAVs naturally add an extra dimension to the visual awareness of the situation. Lastly, UAVs that arrive to the incident site earlier shortens the time to actively engage in the mission.

Now, it seems convenient to take advantage of both components mentioned above by coupling them. However, there are specific limitations that we have to overcome to do so. Firstly, getting immersive media on air requires a stable uplink bitrate from 25 Mbps to 150 Mbps for standard 4K resolution video and 10K resolution video, respectively. Secondly, the transmission stream should maintain low latency, and more stringently, it should be as close to real-time as possible. Although the definition of being real-time is changing as the transmission technologies including radio access techniques and transportation layer protocols are improving, we say the live stream has low latency and ultra low latency when it is getting streamed with sub 5 seconds and 1 second latency, respectively \cite{b4}. 

While maintaining low latency for media transportation is an important focus of this paper, it is also imperative for the entire system to maintain low latency and stability. To do so, the communication latency should be optimized together with the computation latency depending on the resources available to the system. While this framework focuses on the practical aspect, it is nevertheless important to note that it is in line with the analytical mobile edge computing researches on such issues like \cite{b5} and \cite{b6}. 

In Section II, we formulate the problem of immersive real-time aerial monitoring, with the emphasis on the interplay and tradeoff between different units capable of different computation and communication. In Section III, we explain our immersive real-time aerial monitoring framework we have designed. Then, in Section IV we illustrate the current capabilities and limitations of 5G networks in general by experimental results that point to the necessity of Section IV, in which we introduce a real-time bidirectional mission planning scheme. Finally, in Section VI, we demonstrate the implementation of the system in action, then wrap up by pointing to future research directions in Section VII.

\section{Problem Formulation and Contribution}
In this section, we formulate a general public safety problem, and the critical moments until the problem terminates. A general public safety problem consists of the problem site, objects in danger, and the entities involved in solving the problem. In a public safety problem, the problem site has its central location given in 3 dimensional coordinates, i.e. (x,y,z), which is the origin of the problem, whether it is the location fire started, an explosion occurred, or a person fell, and so on. For certain types of incidents, the origin may not be unique, but for simplicity, we consider the origin to be unique. Next we define the range of the problem as a 3 dimensional volume, where the original source of incident is causing a significant change of states to occur, e.g. buildings that caught on fire, explosion radius, or possible gunshot range. Secondly, the objects in danger are individuals and properties within the problem site's range, excluding the source of the problem. Finally, the entities involved in solving the problem are the primary mission executors, e.g. firetruck, fire fighters, and public UAVs, directly involved and exclusively interested in solving the problem.

Now we have to formulate the critical moments until the problem fully terminates. The start time is defined as the time the problem occurs at its origin. The observed time is the time when a first observer, not restricted to a human, notices the problem site or its range. The reported time is the time when an observer issues the problem to a public safety agency, e.g. fire station. The virtual awareness time is the time when an entity from the agency gains direct visual and auditory awareness of the problem site; note this is different from gaining a secondhand awareness from an observer in site. The physical awareness time is the time when an entity from the agency physically arrives at the problem site. The termination time is when the entire range of the problem site is resolved of its issue. The goal of the general problem is to shorten the duration between each critical moments defined above while minimizing the damage incurred to the objects in danger. In Section III, we illustrate an effective framework to reduce the duration between the reported time and virtual awareness time by merging different state-of-the-art components to design an immersive aerial streaming pipeline. Subsequently in Section V, we introduce an effective way to minimize the damage incurred to the objects in danger by leveraging communication and computation tradeoff that arises from our framework in Section III.

In short, this work is motivated by the needs for developing an immersive aerial monitoring system that effectively aids public safety missions. This work also tries to address the difficulty of high data rate aerial communication in the real world. Finally, on top of the low latency capabilities provided by the 5G network, it tries to minimize damage by enabling a stable low latency network for the framework mentioned above by leveraging the trade-off between communication and computation. This paper presents three major contributions to the field of aerial monitoring for public safety: 
\begin{itemize}
\item Practicality: rather than taking a rigorous theoretical approach, this framework takes a very straightforward and practical approach for accomplishing the objectives within the frame of cooperative mission execution for public safety. This includes the efforts to merge different state-of-the-art projects, both open source and proprietary, to achieve the best result possible.
\item Bidirectional Mission Planning: within the frame of planning missions for public safety, instead of taking a centralized mission structure, this framework points toward a distributed mission structure inspired by the extended connectivity promised by the 5G networks, enabling each agents in the mission to participate in the mission more fluidly in real-time.
\item Commercial 5G network: to maintain the sense of practicality to the highest level while trying to satisfy the stringent requirements for low latency, connectivity, and reliability along with wide range of support for diverse missions, this framework uses a commercial 5G network, which is still evolving but equally promising.   
\end{itemize}

\section{Immersive Real-time Aerial Monitoring Framework}

\subsection{Equipment}
Following equipment are sufficient to cover the needs for the framework. Some components may be switched with a different module that has a similar or better capability, but each component's compatibility has to be checked strictly.
\begin{figure}[t]
\centerline{\includegraphics[width=8.7cm]{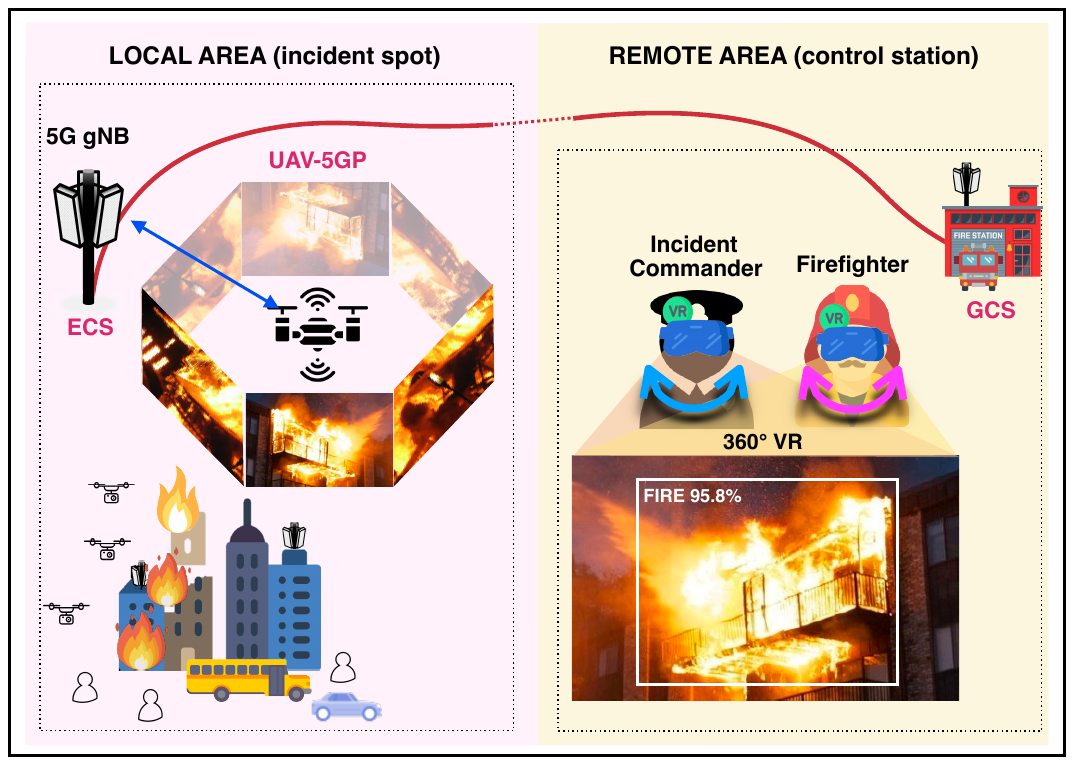}}
\caption{System Equipment: UAV-5GP, in the local area where the incident occurs, is connected to 5G gNB that is connected to the GCS, in a remote area, through 5G core. UAV-5GP mainly transmits high quality immersive media to the users, e.g. fire fighters and fire trucks.}
\label{fig1}
\end{figure}
\begin{itemize}
\item Ground Control Station (GCS): Heavy Computing Server (HCS) and Virtual Reality Head-Mounted Device (VR HMD, Oculus Rift CV1). 
\item Edge Computing Station (ECS): Light Computing Server (LCS) 
\item Unmanned Aerial Vehicle 5G Platform (UAV-5GP): Drone (DJI M600 Pro), 5G device (Samsung Galaxy S10 5G), Onboard Computer (NVidia Jetson Xavier), Immersive Camera (Insta 360 Pro 2) 
\end{itemize}

\begin{figure}[t]
\centerline{\includegraphics[width=7cm]{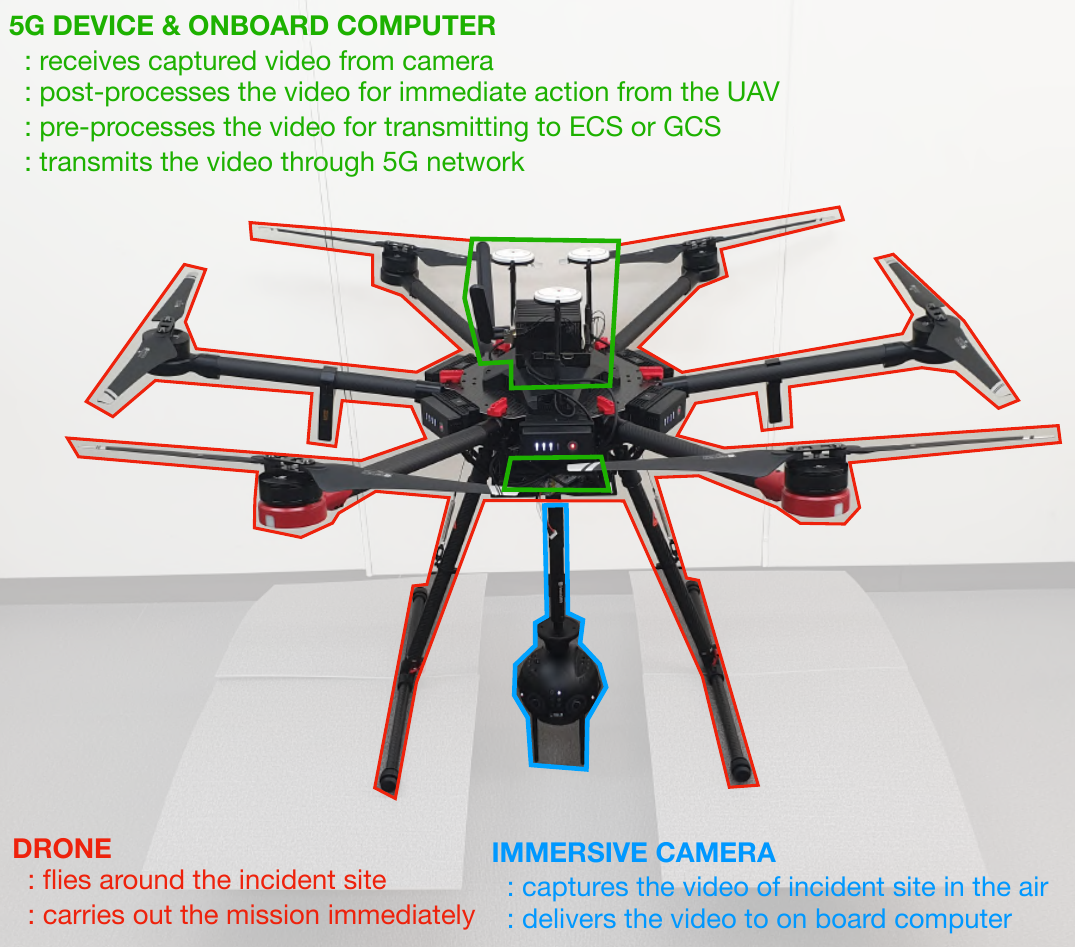}}
\caption{UAV-5GP: the hexarotor drone has an onboard computer, 5G device, and immersive camera mounted on it.}
\label{fig2}
\end{figure}
The GCS, which is terrestrial, indoor, static, and wired, has the highest computing and communicating capabilities. It also has most programs relevant to the mission in it. However, it is the farthest from the problem site, which makes it the entity with highest transportation latency. Some entities located at the GCS are virtually present at the incident site. For virtual presence, Oculus Rift CV1 was used for its compatibility with openHMD as well as its own proprietary libraries. 

The ECS, which is terrestrial, moving, and wireless in a dedicated manner, e.g. moving eNB, has medium computing and communicating capabilities. It has some but not all programs relevant to the mission. Programs that are frequently used are cached in the ECS. It is at somewhere between GCS and the problem site, and virtual presence is possible at the ECS too.

The UAV-5GP, which is aerial, moving, and wireless, has low computing and communicating capabilities. It has the least but most frequently used programs relevant to the mission cached in its hardware. Rotary-wing UAVs are chosen among different types of UAVs to satisfy the flexibility and precision required by the missions. The ability to hover enables monitoring to be more directed, stable, and precise. The major drawback of using a rotary wing is its battery duration, so this factor has to be taken into consideration according to the type of mission the UAVs are involved in. For the experimental purposes, DJI M600 Pro, which is a hexarotor drone, was used as the UAV, and Samsung Galaxy S10 5G was used as the 5G device. Lastly, Insta 360 Pro 2 was attached to capture the situation in 360 videos, and Magewell Pro Capture 4K Plus was mounted on the onboard computer to capture the video stream from the 360 camera.  

Since visual monitoring and processing is required at all levels, all equipment should be capable of certain level of GPU acceleration. Nvidia Jetson Xavier module was mounted on the UAV to enable GPU acceleration; Lenovo Legion Y740-15IRhg acted as the LCS; and Nvidia RTX 2080 ti was mounted on the HCS at the GCS.

\subsection{Real-time High Quality Video Streaming and Processing Pipeline}
Equipment specified above ensures the capabilities to stream and process high quality immersive videos in real-time. The end-to-end latency of the overall pipeline consists of mainly four elements: communication, encoding, decoding, and processing, as it is modeled in the following equation.

The overall pipeline for Real-time High Quality Video Streaming and Processing is shown below. 
\begin{figure}[t]
\centerline{\includegraphics[width=8.7cm]{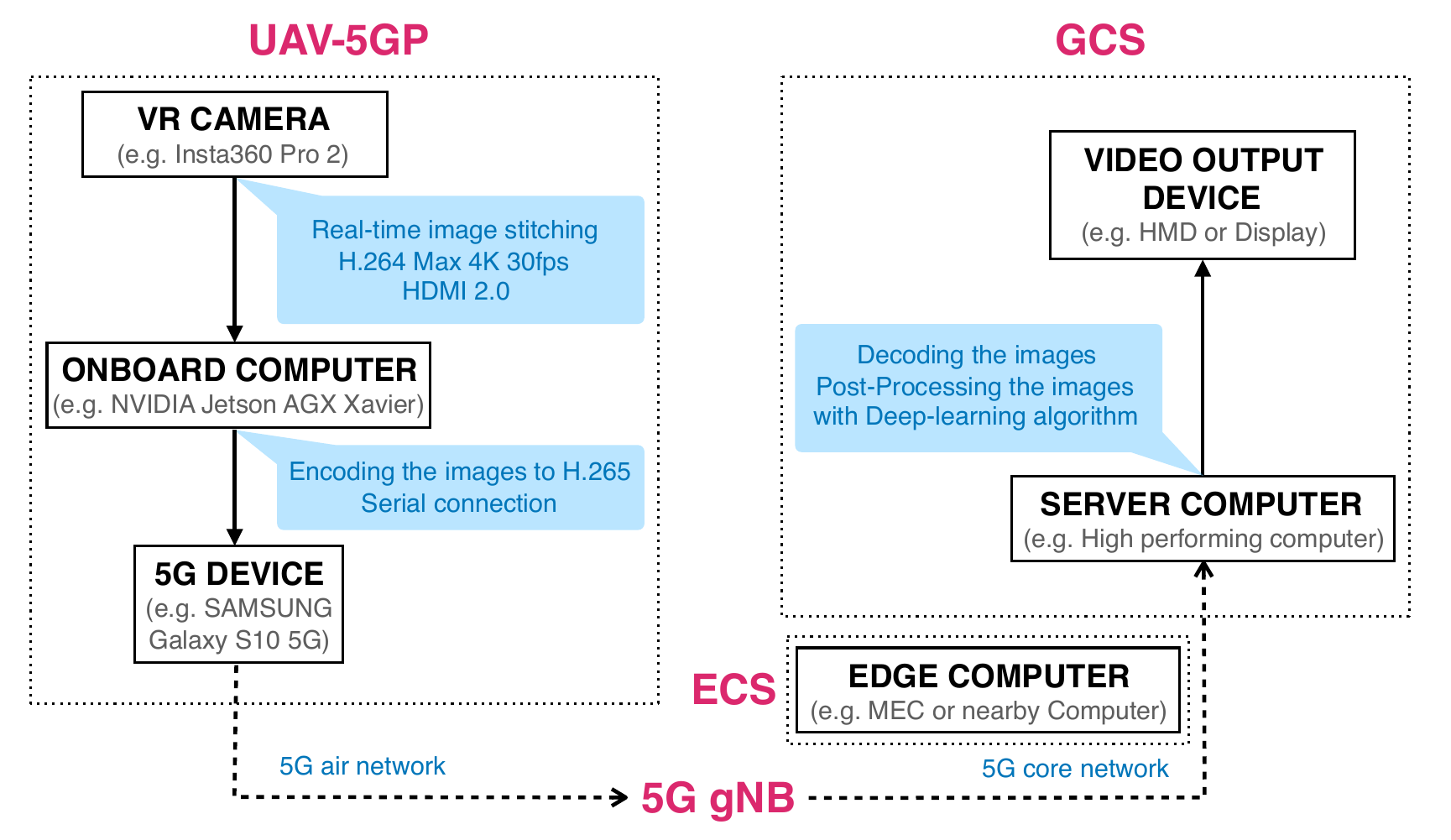}}
\caption{Real-time High Quality Video Streaming and Processing Pipeline.}
\label{fig3}
\end{figure}
\begin{equation}
T_{E2E}=T_{enc}+T_{comm}+T_{dec}+T_{proc}\label{eq}
\end{equation}

\begin{itemize}
    \item $T_{enc}$: As shown in Fig. 3, it covers the following time spent inside the UAV-5GP: stitching captured videos from multiples devices for a VR video, encoding (h264, h265, and etc.) that fits the specific requirements at the ECS or GCS for a more general video.
\item $T_{comm}$: Latency incurred during the transportation of data to ECS or GCS through a 5G gNB. This depends on the network protocols (UDP, TCP, SRT, and etc.) and the communication channel. This latency element is inherently more uncertain than the other elements, thus minimization of $T_{comm}$ is necessary for system stability. This is the main reason behind the design of BIRD scheme introduced in Section V and where the scheme exploits system gain.
\item $T_{dec}$: Latency for decoding the video into a format that the user can read. It depends on the computation power and thus variable.
\item $T_{proc}$: Latency required for post-processing the data received by the user for immediate response to the data to happen. This element is the most dependent on computation power, especially due to the recent developments of programs with neural networks.
\end{itemize}

The length of the pipeline varies depending on where the user acquires and uses the data. For instance, when the GCS uses the data, the data must travel to the GCS physically. In contrary, when the UAV is the source of the data as well as the user, $T_{comm}$ to a server may be a waste. However, the channel conditions may also affect the destination of the data. For example, it is favorable for the GCS to acquire the information when the channel is good and more so when heavy computation is required. Conversely, a bad channel and low complexity makes the UAV-5GP the most likely user, minimizing the latency by omitting $T_{enc}$, $T_{comm}$, and $T_{dec}$.
\begin{figure}[t]
\centerline{\includegraphics[width=8.7cm]{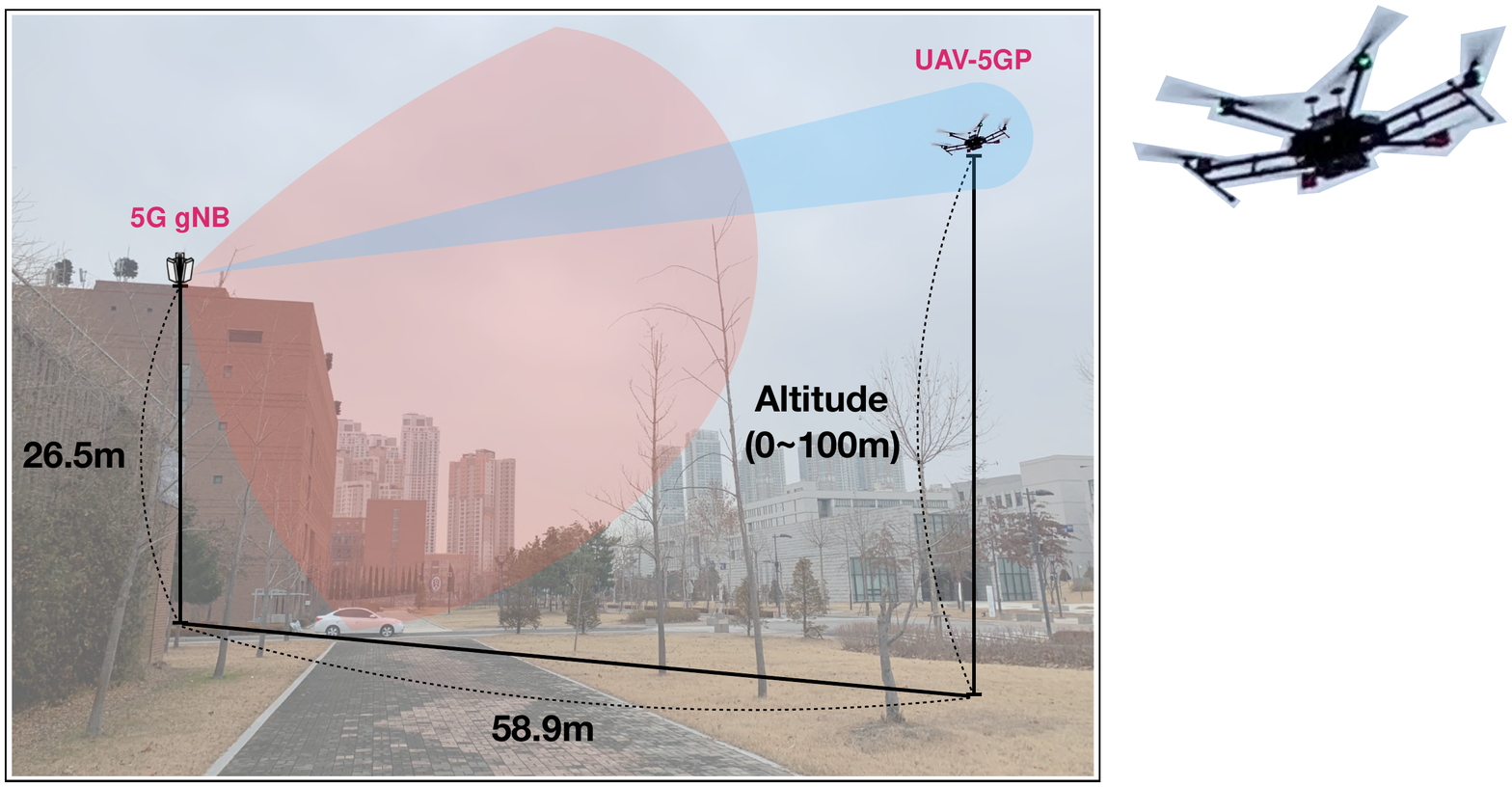}}\vspace{.2in}
\caption{Experimental setup: 5G gNB, which is 26.5 m above ground level, and UAV-5GP, which is varied from 0 to 100 m above ground level, is separated 58.9 m apart from each other at the ground level. }
\label{fig4}
\end{figure}
\section{Aerial 5G Throughput Experiments}
Although the 5G network gifts us with higher data rates, reliability, and lower latency under certain settings, it causes more difficulty in other settings. To test its level of robustness in different settings, especially in settings that are relevant to public safety missions, we provide the following results. 

\subsection{Experiment Setting}\label{AA}
The following results are valid under the current configuration and firmware of various components used during the experiment, some significant changes made to the 5G core and gNB parameters may alter the tendencies shown here. 

A single 5G gNB (32TRX active antenna unit from Samsung Electronics) was located at 37°22'49.4"N 126°40'10.6"E, 26.5 m above ground level. At the time of the experiment, the gNB was non-standalone (NSA), and operated at 3.5GHz. 
The UAV-5GP was located at 37°38'03"N, 126°67'03"E, which is 58.9 m away from the gNB. 
The experiment was conducted during the afternoon hours, and the location was fixed to isolate location specific channel characteristics.
On the UAV-5GP, a 5G device was attached face up on top of the UAV, oriented to the direction of the gNB.  

\subsection{Altitude}\label{AB}
Depending on the mission, drones have to fly at different altitudes. To test how the throughput changes with different altitudes, we measure the throughput at different altitudes. To isolate the effect of altitude, we measured the performance while the drone was in position mode, in which the drone is fixed to a position in all axis of movement. The performance was isolated to measure the instances at positional and motional stability. In addition, network performance was tested with FTP mode to ensure full allocation of resources. Finally, the range of altitudes was chosen deliberately to cover the range of possible public safety missions.

\subsection{Angular velocity}
Along with the effect of flight altitude, the effect of angular velocity is taken into consideration. Significant angular velocity occurs when the drone abruptly changes direction and when it rotates along a specified axis. The UAV was fixed at 10 m and 30 m above the ground, and it rotated with the speed of 0.7757 rad/sec on average. 

\begin{figure}[t]
\centerline{\includegraphics[width=8cm]{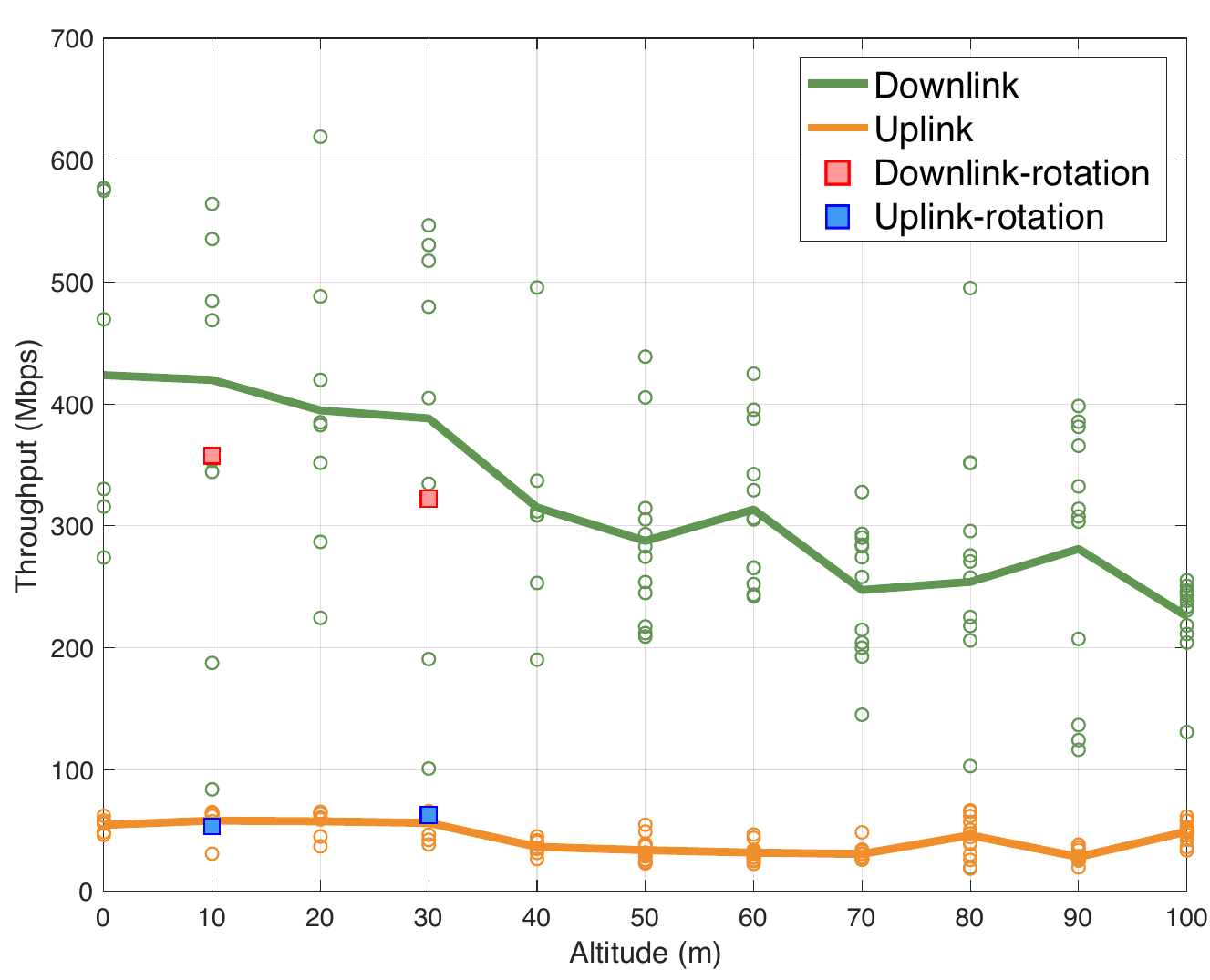}}
\caption{Aerial 5G downlink and uplink throughput experiment results depending on the altitude. The altitude varies from 0 m to 100 m above ground with 10 m interval. For each altitude, at least 3 GB and 500 MB of data was transmitted for downlink and uplink, respectively.  
Additionally, rotation's effect on throughput is also investigated.}
\label{fig5}
\end{figure}

\begin{table}[t]
\caption{Latency and Throughput on air}
\begin{center}
\begin{tabular}{|c|c|c|c|}
\hline
\textbf{Performance}&\multicolumn{3}{|c|}{\textbf{Isolated Flight Aspects}} \\
\cline{2-4} 
\textbf{Metric} & \textbf{\textit{Rotation}}& \textbf{\textit{Low altitude}}& \textbf{\textit{High altitude}} \\
\hline
Latency (ms) & 19.8 & 20.06 & 22.28 \\
\hline
DL Throughput (Mbps) & 339.97 & 356.77 & 264.62 \\
\hline
UL Throughput (Mbps) & 57.99 & 48.13 & 37.12  \\
\hline
\multicolumn{4}{l}{$^{\mathrm{a}}$Low and high altitude denote 0-50 m and 50-100 m respectively.}
\end{tabular}
\label{tab1}
\end{center}
\end{table}

\subsection{Analysis}
In Fig.~\ref{fig5}, the graphs illustrate downlink and uplink throughput performances as the altitude is altered and rotation is applied. The lines represent throughput for the altitude experiment, circular dots represent real sample throughput, and the filled squares represent throughput for the rotation experiment.

As it is shown in Table~1, the network performed robustly against different altitudes within the range of the experiment, more so for uplink communication (with standard deviation of 11.83) than downlink communication (with standard deviation of 72.09). By dividing the range into high and low altitudes, the overall performance degradation is mild but evident, which is in accordance with theory. Here, we provide possible explanations for the performance illustrated by the experiment. First of all, downlink has higher capacity than uplink mainly for these reasons at this moment: downlink is supported by 256 QAM modulation from a 4T4R radio unit whereas uplink from the 5G device we used is supported by only 64 QAM with 1T4R units with limited power. On the other hand, downlink is less robust than uplink because its high level of modulation and coding scheme rapidly degrades as the channel condition weakens. Also, for downlink, MIMO layers decrease from 4 layers opposed to uplink only has one layer to begin with. Finally, rotation had no significant effect on the network performance, which is an expected result for the placement and orientation of the 5G device on the UAV.

The experiment on a real field using commercial 5G network has its own merits but it comes with a cost. Although some elements of flight missions are isolated to test the tendencies of performance, it may not be considered as a definitive indicator of performance due to following sources of error. Firstly, drones are never in precise position mode, but rather in approximate position mode. Secondly, this was only tested in one setting, using the base station located above the power plant building in Yonsei University International Campus, Incheon, Korea. Despite these sources of error, the three experiments provided above are useful because they provide evidence for robust network performance for the possible settings in a public safety mission.

\section{Planning Missions for Public Safety}
In this section, we provide an example case for public safety then introduce a novel system that suits the need for such case.  
\subsection{An Example Mission for Public Safety}
A fire incident breaks out in an urban area. Fire engine equipped with LCS that is connected to 5G network is on its way to the incident site. Before the fire truck arrives, a drone fleet, with each drone connected to 5G network and having its own sensing and computing devices, arrives on the site. Although most commercial drones have limited flight duration, i.e. under 20 minutes, it is sufficient to cover the mission until the arrival of the fire truck; and we define this time as the pre-arrival time. The missions executed during the pre-arrival time is vital for the following reasons. Firstly, by allowing virtual presence in the fire site for fire fighters in the fire truck, fire fighters can obtain accurate information regarding the fire sight that they need. Secondly, fire fighters can command the drone agents in the fire site to execute the missions that the drone agents can handle, e.g. aiding evacuation by providing guidance.     
\subsection{BIdirectional Real-time Distributed (BIRD) Scheme for Autonomous Aerial Fire Fighting}

The main strength of BIRD is its ability to maintain low latency. Due to the nature of the problem, the necessity to maintain low latency is inevitable. Low latency is achieved by striving for autonomy, its reliance on low complexity algorithms, and efficient distribution of workloads.

To elaborate, the distribution happens in two directions: sensing and autonomous decision making. For the UAVs on site to be considered as an autonomous agent, it must possess the ability to decide with a certain level of intelligence. The decision happens in the realm of trajectory planning, which is more traditional, and situation awareness, which is more modern. Consequently, UAVs are not simply agents for sensing, storing, and delivering message, but they act as independent decision makers for certain parts of the mission like having curiosity for more detail, e.g. capturing the video at a better location. However, for safety concerns, level of autonomy is still likely to be controlled by the primary agent, typically the fire fighter inside the fire engine. Secondly, sensing is distributed in a different manner. Distributed sensing has its foundation on distributed intelligence. Because each agent, whether it be the primary or the secondary, possesses the ability to decide for itself, and the agents decide based on the results derived from the sensory data, the data becomes relevant for both types of agents. To add, the UAVs act upon the result by processing their own information autonomously, e.g. taking a better look at a person, and they can also have awareness on what is happening according to the data received from the fire truck. For instance, the human fire fighters may be delayed for the mission, then the drones can be aware of their delay by processing their location and traffic conditions.

Again, low latency should be maintained by leveraging the communication and computation tradeoff described in Section III. Some essential parts for doing so are designing and using low complexity algorithms on the UAV-5GP, deciding which resources to exploit according to the constraints set by the situation and mission specifics.

\begin{figure*}
\centering
    \includegraphics[width=17cm]{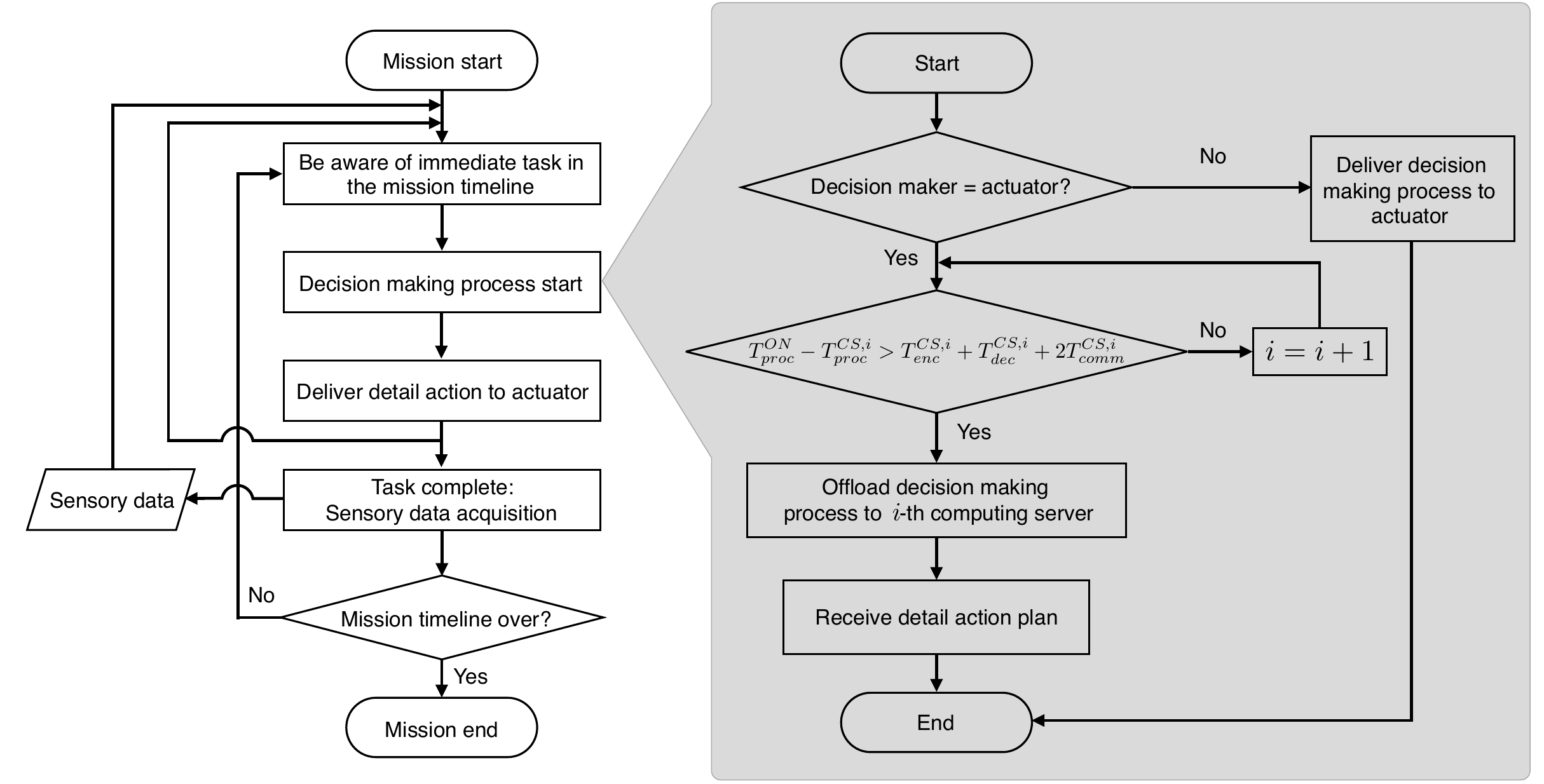}
    \caption{Flow chart of bidirectional real-time distributed scheme for autonomous aerial fire fighting. All entities participate in the mission as candidates for sensing data and planning tasks. Data is transferred in all directions and mission is deployed in a decentralized manner. Simple optimization policy is used for offloading tasks, and mission timeline is updated to ensure situation awareness for all entities.} 
\label{fig8}
\end{figure*}

The technical focus of BIRD is to figure out the current position within the timeline of a mission to realize situation awareness for both autonomous and human fire fighters, so that appropriate actions could be taken to successfully minimize the critical times defined in Section II. First, we define a setting with a UAV-5GP and servers, $S_i = \{i : i = server \ index\}$, where $S_0$ corresponds to the UAV-5GP and the rest of the indices corresponding to the linked computing servers including the GCS and ECS. 

To explain the update procedure, at each time instance $t_i$, specified by an update interval $t_{int}$, the UAV issues an update request to the servers containing the current position within the timeline $t_{pos}$ and the list of programs $p_i$ for a task or part of a task to offload. The list of tasks, e.g. object detection, VR stitching, trajectory optimization, and etc.,  depend on the necessary tasks,  e.g. surveillance, evacuation, and etc., that are explicitly ordered by a commander or implicitly required by the current timeline position. Each task is then matched to programs that exist within program tables, which indicate the programs each server has coupled with their capability and latency information. The matched programs are then offloaded to the most appropriate server by an offloading policy, which is decided by the optimizing equation (\ref{eq}). 

A server responds to the update request by executing the program, returning the result of the program, and reporting its current location (for a moving server). After all update requests are responded, the UAV-5GP becomes fully aware of the situation and updates $t_{pos}$ to the appropriate position. The workflow of BIRD during the pre-arrival time is shown below as a schematic.

\section{Conclusion}
In short, the paper presents a practical framework that uses the recent advances of 5G and UAVs for improving public safety missions. We formulate a general problem structure to cover and assess the improvements brought to public safety with emerging technologies. With the equipment and system configurations presented in the paper, one can implement a system suitable for aerial streaming of high quality immersive media via 5G. After we address the suitability of our framework by providing evidence for aerial 5G network robustness under public safety mission settings, we end by presenting a novel mission planning scheme that could improve missions by enabling bidirectional, real-time, and distributed mission planning. Here, we end the paper with possible directions for further research.

\subsection{Future Work}
Whereas drones were only connected directly to computing servers in this work, drone-to-drone network could be added to further advance efficient and accurate monitoring of the situation. Furthermore, drones could be linked in a hierarchical manner as shown in \cite{b2}.

This work is only concerned with reducing the duration between reported time and virtual awareness time, but further research on sensor networks can reduce the duration between the start time and the reported time. Another line of work that may reduce the critical times by increasing network performance is designing efficient ways to control the orientation and positioning of UAV-5GP, e.g. real-time adjustment of 5G device placement on a UAV-5GP. 
Lastly, experimental results that show the effectiveness of the proposed framework in simulated missions can further strengthen the points made in this paper.

\end{document}